Article

# On-Axis Optical Trapping with Vortex Beams: The Role of the Multipolar Decomposition

Iker Gómez-Viloria,* Álvaro Nodar, Martín Molezuelas-Ferreras, Jorge Olmos-Trigo, Ángel Cifuentes, Miriam Martínez, Miguel Varga, and Gabriel Molina-Terriza*

Cite This: ACS Photonics 2024, 11, 626−633  •  Read Online

ACCESS | 📊 Metrics & More | 📄 Article Recommendations | 🔗 Supporting Information

**ABSTRACT:** Optical trapping is a well-established, decades old technology with applications in several fields of research. The most common scenario deals with particles that tend to be centered on the brightest part of the optical trap. Consequently, the optical forces keep the particle away from the dark zones of the beam. However, this is not the case when a focused doughnut-shaped beam generates on-axis trapping. In this system, the particle is centered on the intensity minima of the laser beam and the bright annular part lies on the periphery of the particle. Researchers have shown great interest in this phenomenon due to its advantage of reducing light interaction with trapped particles and the intriguing increase in the trapping strength. This work presents experimental and theoretical results that extend the analysis of on-axis trapping with light vortex beams. Specifically, in our experiments, we trap micron-sized spherical silica (SiO$_2$) particles in water and we measure, through the power spectrum density method, the trap stiffness constant $\kappa$ generated by vortex beams with different topological charge orders. The optical forces are calculated from the exact solutions of the electromagnetic fields provided by the generalized Lorentz–Mie theory. We show a remarkable agreement between the theoretical prediction and the experimental measurements of $\kappa$. Moreover, our numerical model gives us information about the electromagnetic fields inside the particle, offering valuable insights into the influence of the electromagnetic fields present in the vortex beam trapping scenario.

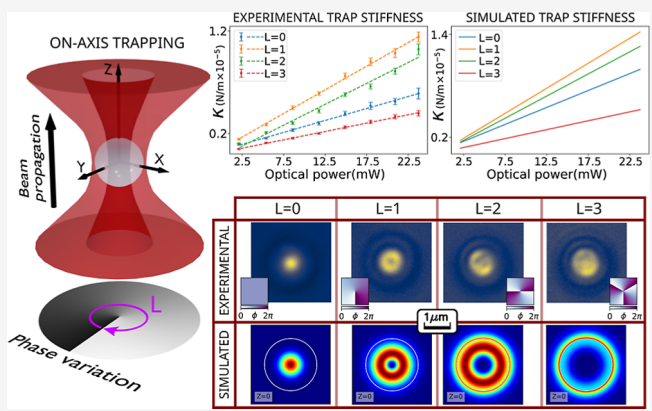

## ■ INTRODUCTION

Since the first experiments showing the possibility of trapping particles using light,[1−3] optical tweezers have been added to the toolbox of several research fields.[4−6] For instance, optical tweezers are used in biology[7] and biochemistry,[8] have allowed the cooling of atoms to ultralow temperatures,[9,10] and contribute to the creation of structured substrates.[11,12] Moreover, in the past few years, optical trapping techniques have been used to levitate small particles to implement fundamental tests of quantum gravity.[13]

Along with the blooming of new applications, there has been significant effort to understand, theoretically and experimentally, the mechanisms behind the optical forces. For a particle of an optical size much smaller than the wavelength of the incident field, the so-called Rayleigh regime, there is a simple and intuitive understanding of how the polarizability of such a particle is related to the optical forces it feels.[14,15] Generally speaking, these particles will be attracted to the spots of high incident beam intensity when their index of refraction is higher than the one of the surrounding medium[16] and repelled when their index of refraction is lower.[17,18] The last scenario, known as "dark-field" trapping, has attracted the interest of the optical trapping community because it involves extremely minimal interaction of the particle with the incident light beam. Indeed, this feature has been exploited in various ways, such as trapping quantum emitters in a configuration where the optical trapping field does not induce fluorescence quenching effects or reducing the temperature increase produced by light absorption in optical traps,[19] which is crucial for many optical tweezers experiments. In fact, depending on the application, high temperatures generated at the focus of optical traps could lead to irreversible damage of the trapped object, especially in biological samples.[20,21] The obvious limiting feature in dark-field trapping systems is the refractive index of the particle that aims to be trapped. In fact, inevitably, this configuration must be dismissed for any optical levitation system, where the medium is air or vacuum. In this regard, there is still an alternative trapping method that is able to highly reduce the amount of light interacting with the particle,[22] specially in its center, and that allows trapping of particles with a higher refractive index than the one of the surrounding medium. This method is based on the use of focused doughnut-shaped beams that generate on-axis trapping of the particle. More precisely, with this technique, the trap presents its stable position when the particle is centered in the incident beam's intensity minimum, while the brightest annular field of it interacts only with the periphery of the particle. Since its initial discovery,[23] on-axis trapping with doughnut-shaped beams quickly attracted the attention of researchers in the optical-trapping community.[24,25] Surprisingly, the very first experiments showed that the trapping strength of certain focused doughnut-shaped beams was even greater than in the bright spot of focused Gaussian beams.[23–26]

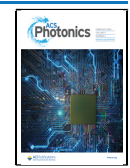

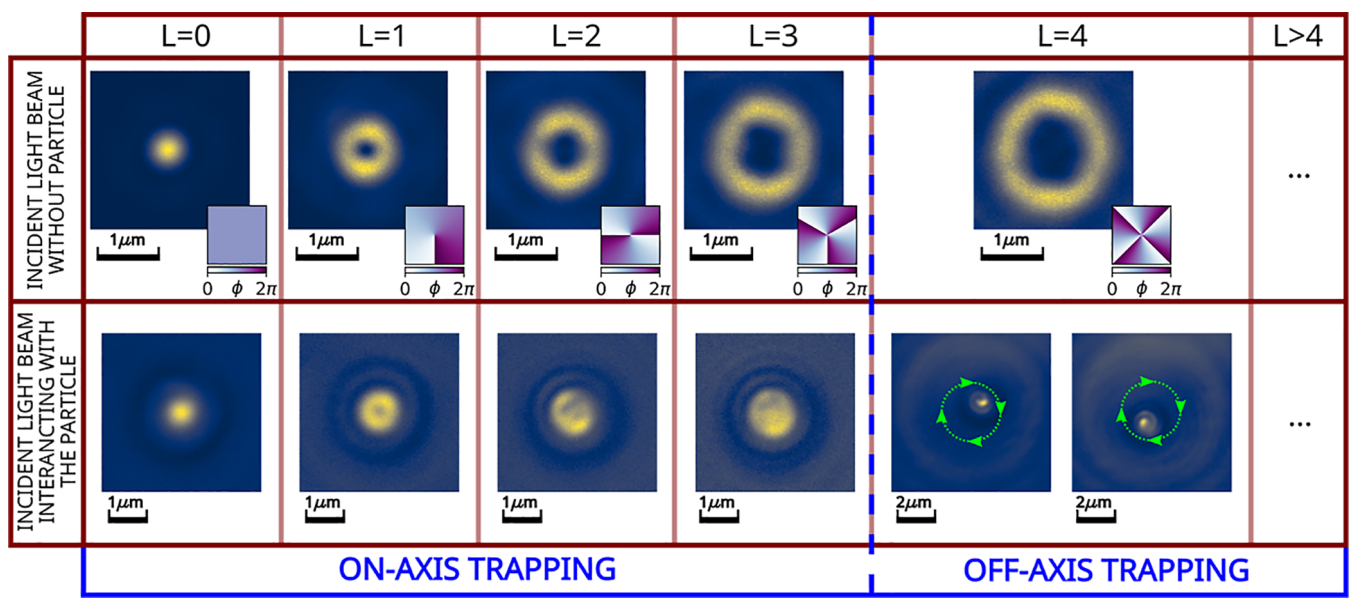

**Figure 1.** Focal plane visualization of trapped silica particles. Illumination of vortex beams with topological charge $L$, focused with an oil-immersion microscope objective with NA = 1.25. An incident light beam without a particle (upper row): the square insets show the spatial phase $\phi$ distribution of each vortex beam, which determines the value of $L$. An incident light beam interacting with the particle (bottom row): we can observe on-axis trapping of a $SiO_2$ spherical particle with a 2 $\mu$m diameter, up to a topological charge $L = 3$. When $L \geq 4$, the particle describes an orbital spinning (off-axis trapping), as it is shown in the two different particle positions of $L = 4$.

In light of this striking phenomenon, there have been several approaches to theoretically describe the on-axis trapping effect with doughnut beams, usually based on the ray-optics model.[26–29] The optical force calculation through the ray-optics model obtains the total linear momentum transferred from the light beam to the particle by discretizing the incident light beam into a bundle of light rays. However, it is well-known that the ray-optics model fails for particles with radius, $a$, close to the wavelength, $\lambda$, of the incident light beam. Therefore, an alternative description of the optical forces is required in such cases. In this regard, generalized Lorentz–Mie theory (GLMT)[30] provides an exact solution to the scattering problem of spherical particles under general illumination conditions, making it the most reliable procedure for calculating the electromagnetic fields contributing to the optical forces.[31–33]

In this work, we present both experimental and theoretical results analyzing the on-axis trapping strength caused by vortex beams over particles in the Mie regime ($\lambda \simeq a$). We model the trapping using the exact GLMT in order to compare it with the experimental results. This allows us to go beyond the usual approach using approximate theories such as ray optics. Moreover, our theoretical approach gives us access to capture the electromagnetic fields inside the particle. In addition to this, we experimentally measure the stiffness constant of the optical trap with vortex beams of four different topological charge orders, including a Gaussian beam. The theoretical results agree qualitatively with the experimental results and show that on-axis traps generated by vortex beams can have a higher stiffness constant than when impinging with a Gaussian beam. On the other hand, the plots of the electric field inside the silica particle confirm that on-axis trapping of doughnut-shaped beams reduces the light impact caused by the incident light beam. Both effects can be understood due to the fact that structured fields, such as Laguerre–Gaussian beams, suppress the dipolar contribution of particles,[34] which are the ones that dominate the scattering force and provide the highest intensity values to the center of the internal electromagnetic fields of the particle.

## ■ ON-AXIS OPTICAL TRAPPING GENERATED BY VORTEX BEAMS IN TERMS OF THEIR MULTIPOLAR DECOMPOSITION

The key element for the appearance of the stable centered (on-axis) optical trapping regime when vortex beams are employed is that large particles support a plethora of multipolar modes.[35–37] Importantly, the coupling of these modes strongly depends on its optical size, which can be characterized by the dimensionless parameter $v = mka$, where $m$ is the refraction index contrast between the particle and the surrounding medium, $a$ is the radius of the particle, and $k = 2\pi/\lambda$ is the wavevector of the incident field. In general, $v$ is proportional to the number of modes that the particle can support. On the other hand, structured beams, such as Laguerre–Gaussian which means that any multipolar order lower than the corresponding value of $|m_z|$ is suppressed if the optical system is cylindrically symmetric[34] (see Supporting Information). This effect is present in the on-axis trapping generated by Laguerre−Gaussian modes with $L = 0$, 1, 2, and 3 that can be observed in Figure 1. In these four cases, despite the suppression of the lowest multipolar modes, the sphere is still effectively coupled to the beam due to its optical size. It is important to remark on the consequences of the suppression of multipolar modes, bounded by the value of $|m_z|$, that occurs in the on-axis configuration. This suppression implies that a polarization swap in the incoming light beam ($p \rightarrow -p$) would also impact its multipolar content and, consequently, the optical forces exerted on the particle.[38] However, due to the mirror symmetry of the spherical object, this effect will be observed only in beams with $L$ being different than zero. In contrast to the on-axis trapping cases, in Figure 1 can be seen that when the mode is too high (in this case, $L = 4$), the particle is no longer stably trapped in the central position and moves to the bright ring (off-axis trapping), similar to what is expected with a small particle in the dipolar regime. In this scenario, when the particle is positioned on the axis, the beam with $L = 4$ suppresses too many lower modes. In other words, the optical size of the particle is insufficient for effective coupling with the lowest available multipolar modes of the beam. In Figure 2, simulations of the multipolar coefficients present in the different experimental trapping situations of Figure 1 are shown only for the on-axis configuration. Here, the decoupling effect produced when the $L$ is increasing, which, for $L = 4$, is able to definitely spoil the stability of the trap in the on-axis configuration. Only the off-axis configuration, where the cylindrical symmetry of the system is broken, would cause the appearance of the lowest multipolar modes,[39] increasing its coupling with the beam. The particle achieves the most effective coupling with the new multipolar modes at the brightest annular part of the beam, resulting in trapping at that position. Moreover, as the trapping field carries orbital angular momentum, the particle starts rotating in the ring in stable albeit nonstationary dynamics.[40−42]

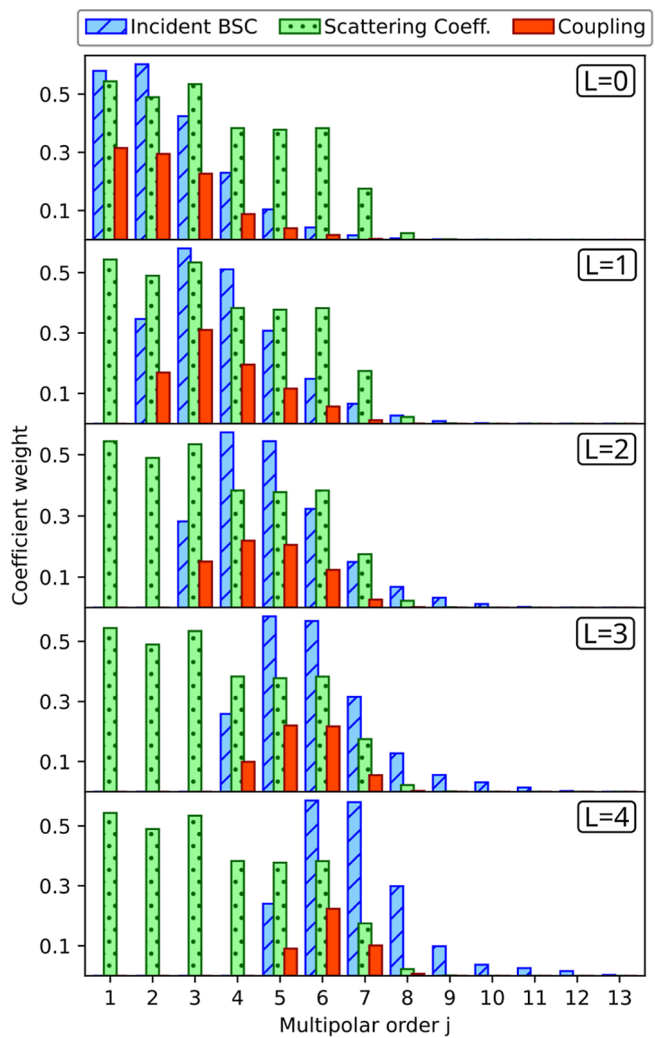

## ■ EXPERIMENTAL MEASUREMENT OF THE TRAP STIFFNESS CONSTANT

In Figure 3, we show the experimental setup used in this work. Here, one can observe that structured light beams are created with a spatial light modulator (SLM). Then, the modulated beam evolves along a 4f system, which reconstructs the beam's SLM-plane wavefront onto the back aperture of an oil-immersion microscope objective with NA = 1.25. This procedure ensures a constant filling of the back aperture of the objective, while the different Laguerre−Gaussian modes are projected. The objective lens employed can tightly focus the different modes of the beam, generating high light intensity gradients at its focus and thus creating an excellent environment for optical trapping. Besides the imaging system used to monitor the stability of the optical traps, the optical setup was also prepared to measure the displacement of the particles with a four-quadrant photodetector. This information can be processed to calculate the stiffness constant $\kappa$ of the trap. To that purpose, we employ the power spectrum density (PSD) method[43,44] (see Supporting Information). This process obtains the oscillation frequencies of the particle from its position measurement through a Fourier transform. Even though our traps are nonstandard, given their stability, modes, can couple more effectively to the higher-order modes and avoid coupling to the lower-order modes, such as the dipolar modes. Figure 1 shows an example of such stable trapping. In this case, we have a silica ($SiO_2$) spherical particle of diameter $d \simeq 2$ $\mu$m trapped by different structured beams in water. The refraction index contrast is $m \simeq 1.1$ and the wavelength we used is $\lambda = 976$ nm, resulting in $\nu \simeq 7$, well above the dipolar regime. The Laguerre−Gaussian or vortex beams employed in Figure 1 are characterized by their topological charge number $L$, i.e., the number of $2\pi$ times the spatial phase of the beam's wavefront winds around its center (see the insets of the upper row of Figure 1). Additionally, the beams we used have left circular polarization (helicity $p = +1$) and, as long as the particle is in an on-axis position, preserve the cylindrical symmetry of the system. This kind of beam has a well-defined total angular momentum in the $z$ direction ($m_z$) with a sharp value of $m_z = L + p = L + 1$, the trapping potential must present a minimum and can be described as a first approximation of a harmonic potential. Therefore, the dynamics of the particle can be modeled as a harmonic oscillator in an overdamped regime. Then, the $\kappa$ parameter can be obtained by fitting the experimental power spectrum as a Lorentzian function. The next step is to extract the value of the corner frequency $f_c$, which is the frequency at which the power of the samples starts to decay. Knowing the value of this parameter, we can use the relation $\kappa = 2\pi\beta f_c$, where $\beta = 6\pi\nu a$ is the friction coefficient, $\nu$ is the viscosity coefficient of the medium, and $a$ is the radius of the particle. In Figure 4, we show the PSD measurement process of $\kappa_x$ and $\kappa_y$, for a silica particle trapped with a focused vortex beam of topological order $L = 2$ and an optical power of 17.5 mW. Note that this is an accurate model of the optical trap as far as the movement of the center of mass of the particle is small and respects the harmonic approximation, otherwise one should include anharmonicities of the trap in its description.[45] In consequence, this method fails to accurately describe the nonstationary trap of the $L = 4$ beam. On the other hand, it allows us to compare theoretically and experimentally the strengths of the different stationary traps produced with structured beams. It is noteworthy that the PSD method is considered one of the most precise techniques in the derivation of the $\kappa$ parameter, essentially due to its capability of filtering the undesired frequencies produced by external sources.[46]

**Figure 2.** Simulations of the squared moduli of the multipolar coefficients present in each experimental trapping beam in the case of the on-axis configuration. The beam shape coefficients (BSC) of the incident light beam (blue-striped bars) are calculated for vortex beams with topological charge $L$, focused by a lens with NA = 1.25. The scattering coefficients (green dotted bars) are formed combining the Mie coefficients of a $SiO_2$ spherical particle with a 2 $\mu$m diameter, surrounded by water, which results in an effective refractive index contrast of $m \simeq 1.1$. The coupling coefficients (red plain bars) are the multiplication of the BSCs and scattering coefficients.

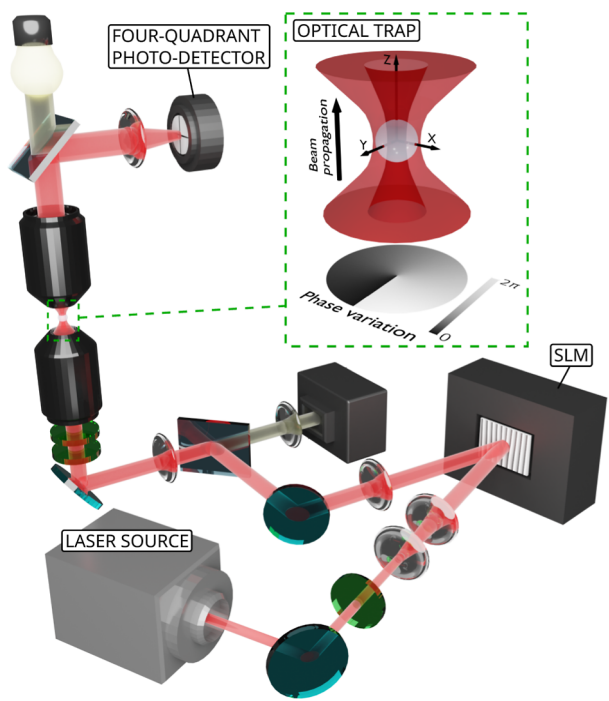

**Figure 3.** Experimental optical tweezers setup. The main subsystems contained in the setup are the wavefront phase modulation system, the imaging system of the *xy*-plane of the trapping region, and the trap stiffness constant ($\kappa$) measurement system. For a detailed description see the Supporting Information.

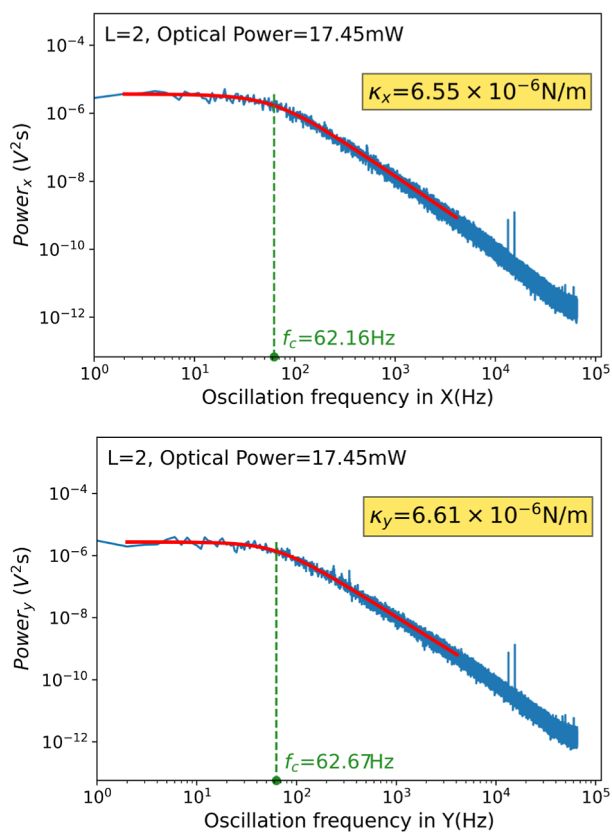

**Figure 4.** Example of the PSD procedure for stiffness constant calculation. Examples of $\kappa_x$ and $\kappa_y$ measurements for an incident light beam with 17.45 mW of optical power and topological charge $L = 2$. The processing of the particle's oscillation frequencies in order to extract the $\kappa$ parameter can be observed, first by applying the Lorentzian fitting and then by determining $f_c$.

## ■ RESULTS AND DISCUSSION

The experimental measurements of the transversal stiffness constant $\kappa_{x/y}$ (averaged value between $\kappa_x$ and $\kappa_y$) are shown in Figure 5a for eight different optical powers ranging from 3.5 to 22.5 mW. We present our results for the modes able to generate on-axis trapping: $L = 0$, 1, 2, and 3. The stiffness constant shows a linear dependence with the optical power for all modes, i.e., $\kappa_{x/y} = \alpha_L P$, where $P$ is the optical power of the trapping beam. $\alpha_L$ allows us to describe the force of the optical trap independently of its power. The linear scaling of the power is a consequence of the weight of the particle being negligible compared to the optical forces in the vertical ($z$) direction. Otherwise, changes in the optical power would produce a variation of the equilibrium position in the $z$-axis, and in consequence, the scaling of $\kappa_{x/y}$ would not be linear. Now, we can compare the trapping strength of the setup when using different modes. A remarkable fact that can be extracted from the experimental figure is that the trapping stiffness significantly increases from $L = 0$ to $L = 1$, for all powers, i.e., $\alpha_0 = 2.6 \pm 0.3 \times 10^{-7}$ (N/m)/mW < $\alpha_1 = 5.1 \pm 0.4 \times 10^{-7}$ (N/m)/mW. The bottom line is that the optical force exerted on the particle is stronger when trapping with a Laguerre—Gaussian mode of $L = 1$ than the one used in typical bright-in-the-center Gaussian beams. This effect seems counterintuitive, given that the equilibrium point of the silica particle is placed in the darkest part of the beam, where the intensity is zero. However, as mentioned above, it has already been observed in previous works of optical trapping with doughnut-shaped beams.[23–26] Our experimental measurements confirm this phenomenon, as can be observed in the curves of Figure 5a, which also shows that $\alpha_2$ is larger than $\alpha_0$, while we have to go to $L = 3$ to notice a decrease in the strength of the trap. These effects can be attributed to the fact that vortex beams suppress the dipolar modes,[34] which are the ones that dominate the scattering force. Thus, the influence of the gradient force is enhanced, which improves the trapping of the sphere.

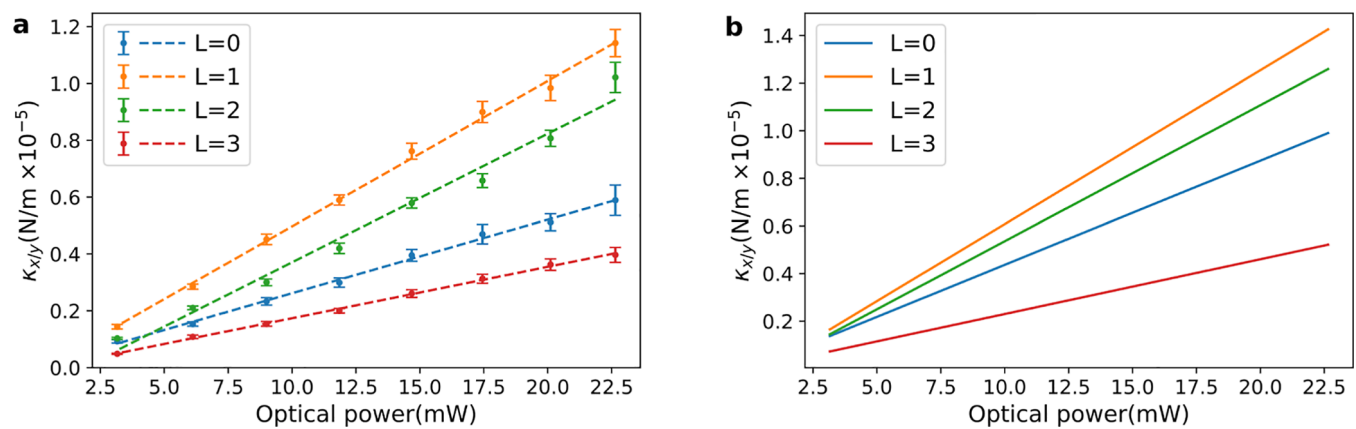

**Figure 5.** Experimental and theoretical transversal trap stiffness constant ($\kappa_{x/y}$) calculation. (a) Experimental $\kappa_{x/y}$ values for vortex beams of topological charges $L = 0$, 1, 2, and 3 as a function of the optical power of the trapping laser beam. The colored dots represent the average value between $\kappa_x$ and $\kappa_y$ measurements for each of the 4 topological charge orders and 8 different optical powers. The dashed lines linearly fit the experimental results for each topological charge order of the vortex beam employed. (b) Theoretical $\kappa_{x/y}$ values for vortex beams of topological charges $L = 0$, 1, 2, and 3 as a function of the optical power of the trapping laser beam.

In order to gain further insights into the physical mechanism behind these effects, we performed a series of numerical simulations of the optical forces in this regime. The simulations are done using the GLMT,[30,47–49] which allows us to decompose the structured light beams into multipolar modes[50–52] and then calculate the scattering properties of the spherical particles in order to compute the force acting on the particle[31–33] (see Supporting Information). With this numerical analysis, it is possible to extract the $\kappa_{x/y}$ parameter for the different modes and optical powers of Figure 5a. The theoretical calculations are shown in Figure 5b, which show excellent qualitative agreement with the experimental results. The model captures the main features of the phenomenon: stability of the particle in the on-axis configuration for modes of $L = 1$, 2, and 3, stronger optical forces for some higher-order modes compared with Gaussian beam trapping, and even the hierarchy of the strengths of the forces of the different modes is replicated. On the other hand, the quantitative values can differ from the experimental ones due to small differences in the diameter of the particle, aberrations remaining in the experimental modes,[53,54] or small differences in their shape not captured by our model. In fact, most of the quantitative disagreement disappears when we consider a global reduction of 15% in the optical power transmitted to the sample. This reduction can be simply attributed to the losses caused by the transmission of light through the planar interfaces present in an oil-immersion objective.[55]

The numerical model allows us to better understand the forces in the three directions or, equivalently, the optical potential traps. The visualization of the electromagnetic field inside the particle also provides valuable insights into the behavior of the optical trap. In Figure 6, we depict the modulus of the electromagnetic field in the on-axis configuration and also the optical force and optical potentials when the beam focus is displaced in $z$ and $x$ directions for different vortex beams. The mathematical representation of displaced electromagnetic fields was obtained following the procedure described in ref 39. In this reference, it is shown how the multipolar content of the incident field could be reorganized in order to generate displaced versions of it. In Figure 6, it is noticeable that, as expected, the field inside the particle is significantly diminished when the incident field mode order $L$ increases. In particular, for $L = 3$, the field at the core of the particle is basically absent. The suppression of the dipolar contribution caused by vortex beams[34] is once again the reason for this phenomenon. The dipolar mode of the incident beam, in contrast to higher modes, is characterized by a high electromagnetic field intensity at the center of the symmetry of the beam. This high intensity field extends into the inner part of the particle. If the dipolar mode is suppressed, then its great central contribution to the internal electromagnetic fields of the particle would be eliminated. Nevertheless, these four cases present a well-defined minimum of potential in the trap. With $L = 3$, a slight shouldering effect is present near the center of the potential well, which forewarns the possibility of a structural change in the potential well as $L$ increases. This is actually confirmed in Figure 7, where the centered stable and stationary trapping regime is not reached for $L = 4$. This finding is also in perfect agreement with the experimental observations shown in Figure 1. In fact, taking a look at the energy values in the transition from $L \leq 3$ to $L > 3$, we can easily notice how the system passes from a unique to a double potential well, which represents passing from the stable centered trapping regime to the nonstationary trapping in the ring of the vortex beam.

## ■ CONCLUSIONS

In conclusion, we have presented a comprehensive analysis of the on-axis trapping effect using dielectric silica particles with a radius similar to the wavelength of the trapping laser in an aqueous medium and light vortex beams. The experimental and theoretical data presented here show that the stiffness constant of the trap can be significantly enhanced when focused vortex beams are employed instead of the widely used Gaussian beams. We have provided full theoretical optical force calculations, showing the appearance of this phenomenon. In fact, our force calculation model accurately predicts the observed trapping strength hierarchy for the different topological charge orders used in the experiment. Additionally, we conducted numerical simulations of the electromagnetic fields of the trap, which show that using higher-order beams minimizes the optical field at the core of the particle while maintaining a stable and stationary centered optical trapping. We have also shown that the suppression of the lowest multipolar modes caused by the use of vortex beams plays an essential role in the description of the optical forces and electromagnetic fields present in these types of optical traps.

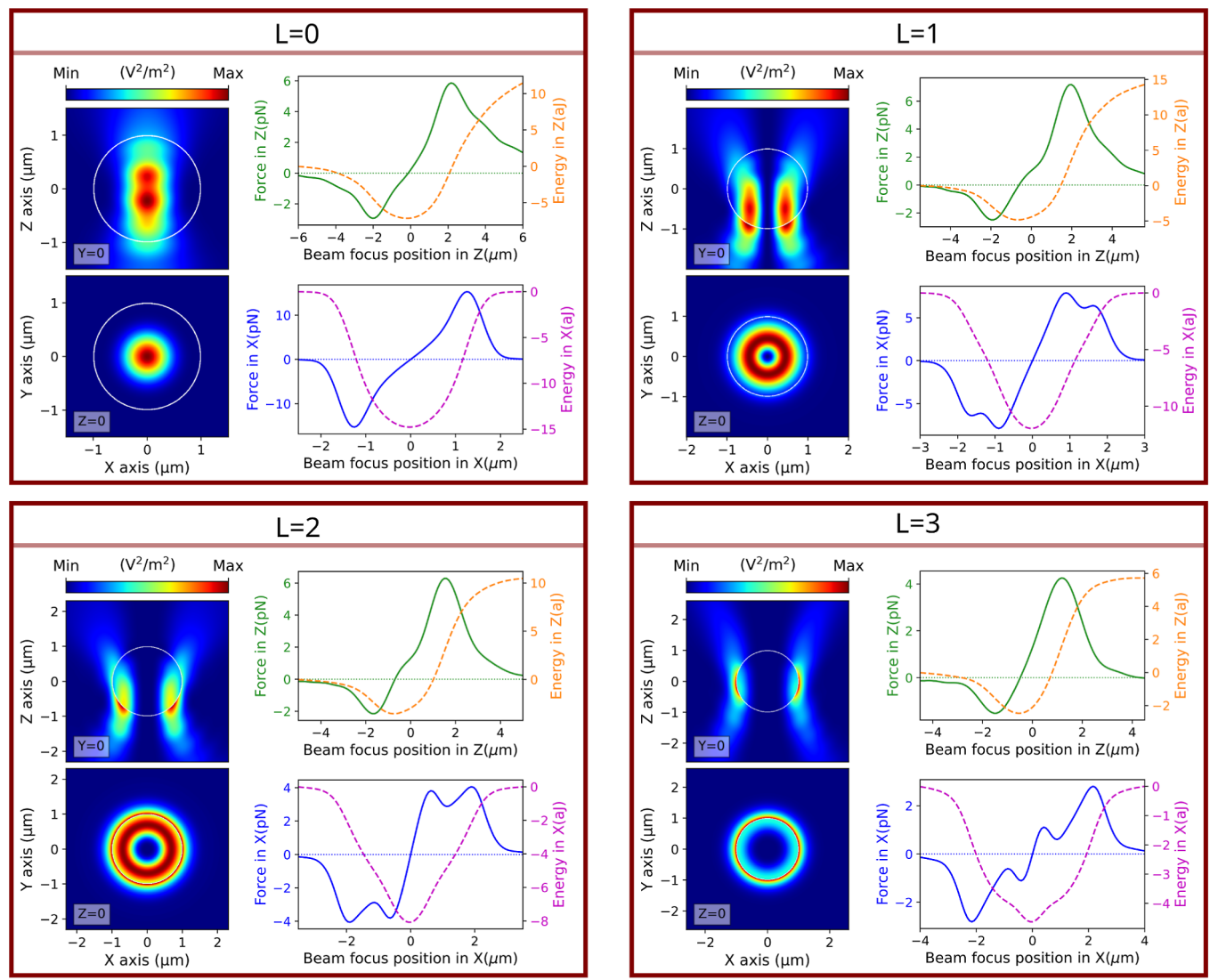

**Figure 6.** Total electric field intensity and optical force and energy simulations of the experimental on-axis trapping cases. Each subfigure represents one of the 4 topological charge orders $L = 0$, 1, 2, and 3 capable to generate on-axis trapping. On the left side of each subfigure: colormaps of the total electric field intensities, at the equilibrium point of each optical trap, in the $xz$- and $xy$-plane. On the right side of each subfigure: 2D plots of the optical force in $z$ (solid green) and its trap energy (dashed orange) and of the optical force in $x$ (solid blue) and its trap energy (dashed purple). The diameter of the silica particle is 2 $\mu$m. The objective-lens NA = 1.25, and the incident wavelength is $\lambda = 976$ nm for all topological charge orders $L$.

## ASSOCIATED CONTENT

### ⓢ Supporting Information

The Supporting Information is available free of charge at https://pubs.acs.org/doi/10.1021/acsphotonics.3c01499.

Theoretical description of electromagnetic fields and forces using the multipolar decomposition of GLMT and the surface integration of the Maxwell stress tensor ($T_{ij}$), description of the experimental setup detailing the wavefront phase modulation system, the imaging system of the trapping region, and the trap stiffness constant

($\kappa_{x/y}$) measurement system, and trap stiffness measurements through the PSD method: operation of the four-quadrant photodetector and Lorentzian fitting procedure of the experimental power spectrum (PDF)

## ■ AUTHOR INFORMATION

### Corresponding Authors

**Iker Gómez-Viloria** − *Centro de Física de Materiales (CFM), CSIC-UPV/EHU, 20018 Donostia-San Sebastian, Spain;* ● orcid.org/0000-0002-0501-5215; Email: iker_gomez@hotmail.com

**Gabriel Molina-Terriza** − *Centro de Física de Materiales (CFM), CSIC-UPV/EHU, 20018 Donostia-San Sebastian, Spain; Donostia International Physics Center, 20018 Donostia-San Sebastian, Spain; IKERBASQUE, Basque Foundation for Science, 48013 Bilbao, Spain;* Email: gabriel.molina.terriza@gmail.com

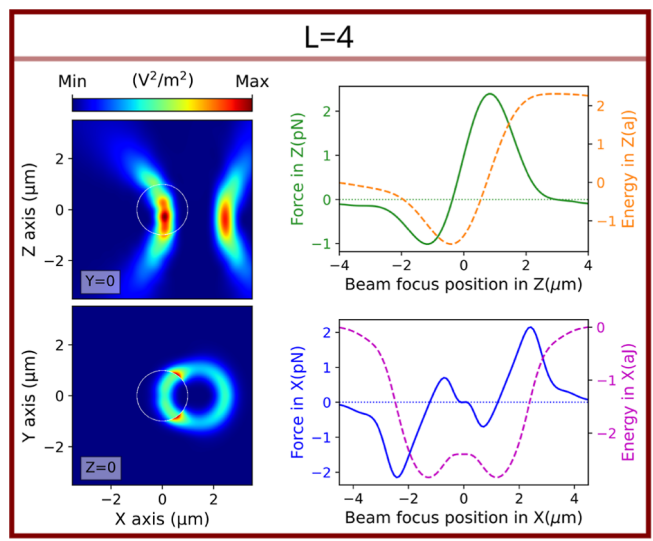

**Figure 7.** Total electric field intensity and optical force and energy simulations of the experimental off-axis trapping case with $L = 4$. On the left side: colormaps of the total electric field intensities, at the equilibrium point the optical trap, in the $xz$- and $xy$-plane. On the right side: 2D plots of the optical force in $z$ (solid green) and its trap energy (dashed orange) and of the optical force in $x$ (solid blue) and its trap energy (dashed purple). The objective-lens NA = 1.25, and the incident wavelength is $\lambda = 976$ nm. Under these experimental conditions, the topological charge $L = 4$ is the first one that cannot generate on-axis trapping for a silica spherical particle with a 2 $\mu$m diameter.

## Authors

**Álvaro Nodar** − *Centro de Fisica de Materiales (CFM), CSIC-UPV/EHU, 20018 Donostia-San Sebastian, Spain*

**Martín Molezuelas-Ferreras** − *Centro de Fisica de Materiales (CFM), CSIC-UPV/EHU, 20018 Donostia-San Sebastian, Spain*

**Jorge Olmos-Trigo** − *Centro de Fisica de Materiales (CFM), CSIC-UPV/EHU, 20018 Donostia-San Sebastian, Spain;* orcid.org/0000-0003-2953-2433

**Ángel Cifuentes** − *Centro de Fisica de Materiales (CFM), CSIC-UPV/EHU, 20018 Donostia-San Sebastian, Spain*

**Miriam Martínez** − *Centro de Fisica de Materiales (CFM), CSIC-UPV/EHU, 20018 Donostia-San Sebastian, Spain*

**Miguel Varga** − *Centro de Fisica de Materiales (CFM), CSIC-UPV/EHU, 20018 Donostia-San Sebastian, Spain; Donostia International Physics Center, 20018 Donostia-San Sebastian, Spain*

Complete contact information is available at:
https://pubs.acs.org/10.1021/acsphotonics.3c01499

## Funding

Iker Gómez-Viloria, Gabriel Molina-Terriza, Miguel Varga, Ángel Cifuentes, and Miriam Martínez acknowledge support from the CSIC Research Platform on Quantum Technologies PTI-001, from IKUR Strategy under the collaboration agreement between Ikerbasque Foundation and DIPC/MPC on behalf of the Department of Education of the Basque Government and from the projects EQC2018-004060-P and PID2022-143268NB-I00 of Ministerio de Ciencia, Innovación y Universidades. Álvaro Nodar acknowledges the financial support from the Spanish Ministry of Science and Innovation and the Spanish government agency of research MCIN/AEI/10.13039/501100011033 through project ref. no. PID2019-107432GB-100 and from the Department of Education, Research and Universities of the Basque Government through project ref. no. IT 1526-22. Martín Molezuelas-Ferreras acknowledges financial support from the Spanish Ministerio de Ciencia, Innovación y Universidades through the FPI PhD fellowship FIS-2017-87363-P. Jorge Olmos-Trigo acknowledges support from the Juan de la Cierva fellowship no. FJC2021-047090-I of MCIN/AEI/10.13039/501100011033 and NextGenerationEU/PRTR.

## Notes

The authors declare no competing financial interest.